\documentclass[aps,preprint,unsortedaddress,floats]{revtex4}
\usepackage{amsmath,amsfonts,graphicx,color}

\begin{document}
\bibliographystyle{revtex}

\textwidth 16cm \textheight 23cm \topmargin -1cm \oddsidemargin
0cm \evensidemargin 0cm

\title{Transient growth in stable collisionless plasma}
\author{Enrico Camporeale$^{a}$}\email{e.camporeale@qmul.ac.uk}
\author{David Burgess$^{a}$}\email[]{d.burgess@qmul.ac.uk}
\author{Thierry Passot$^{b}$}\email[]{passsot@oca.eu}

\affiliation{a) Queen Mary University of London, Mile End Road,
London E1 4NS, UK} \affiliation{b) University of Nice Sophia
Antipolis, CNRS, Observatoire de la C\^{o}te d'Azur, B.P. 4229,
06304 Nice Cedex 4, France}

\date{\today}

\begin{abstract}
The first kinetic study of transient growth for a collisionless
homogeneous Maxwellian plasma in a uniform magnetic field is
presented. A system which is linearly stable may display transient
growth if the linear operator describing its evolution is
non-normal, so that its eigenvectors are non-orthogonal. In order
to include plasma kinetic effects a Landau fluid model is
employed. The linear operator of the model is shown to be
non-normal and the results suggest that the nonnormality of a
collisionless plasma is intrinsically related to its kinetic
nature, with the transient growth being more accentuated for
smaller scales and higher plasma beta. The results based on linear
spectral theory have been confirmed with nonlinear simulations.
\end{abstract}

\pacs{}
%\keywords{}

\maketitle

\begin{figure}
\includegraphics[width=18pc]{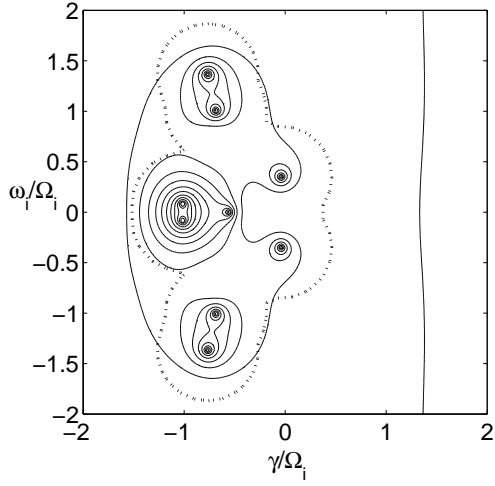}
\caption{Contours of the $\varepsilon$-pseudospectrum for a plasma with $\beta=10$ and $k=1$. Contours are plotted for $\log_{10}\varepsilon=-4.9,-4.7,\ldots,-2.7$. The dotted line is how the $\varepsilon$-contour would appear if the operator were normal, for $\varepsilon=0.5$}\label{contour_inset}
\end{figure}

\begin{figure}
\includegraphics[width=18pc]{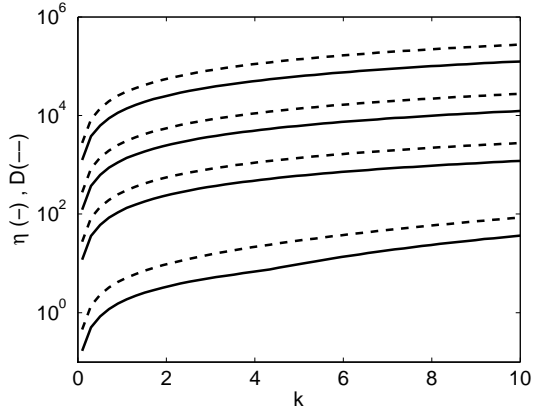}
\caption{Numerical abscissa $\eta$ (solid) and departure from
normality $D$ (dashed), as a function of $k$ (normalized to the
ion Larmor radius). The four curves are for $\beta=0.1,1,5,10$
from below to above.}\label{num_abs-dep-k}
\end{figure}

\begin{figure}
\includegraphics[width=18pc]{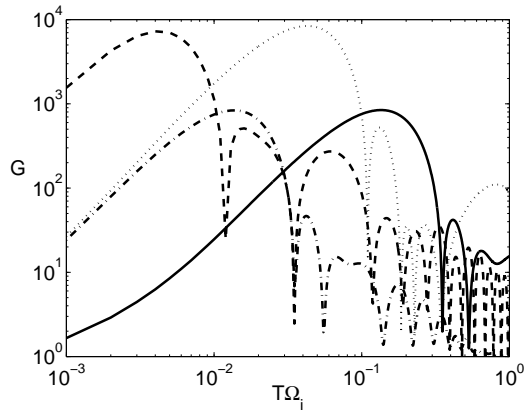}
\caption{Amplification $G$ as a function of time $T$ (normalized to the ion cyclotron frequency $\Omega_i$), as computed via a nonlinear Landau fluid code. The initial perturbation is the one that maximizes the amplification. The four curves are for the following parameters: $\beta=1,k=1$ (solid line); $\beta=10,k=1$ (dotted); $\beta=1,k=10$ (dash-dotted); $\beta=10,k=10$ (dashed)}\label{sim}
\end{figure}

\begin{figure}
\includegraphics[width=18pc]{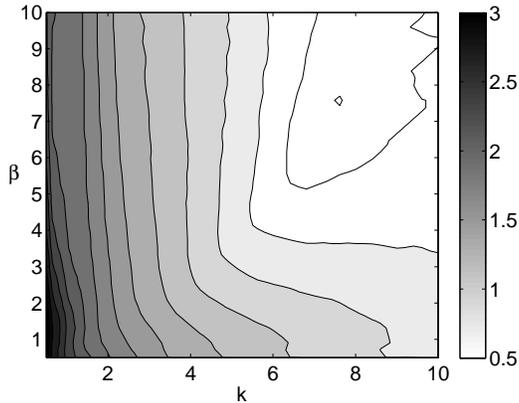}
\caption{Contour plot of time $\Theta$ for which $\|e^{\mathbf{A}\Theta}\|\leq 1$, that is the time for which transient growth is exhausted, as a function of $k$ and $\beta$. Colorbar is given in $\log_{10}\Theta$, and normalized to the ion cyclotron frequency $\Omega_i$.}\label{contour}
\end{figure}

%\section{Introduction}
The stability theory of hot magnetized plasma has been
historically based on normal mode analysis, which has led over
time to the identification of a great variety of instabilities and
waves. The normal mode approach is usually applied to the linear
approximation of the dynamical equations of the system, where a
small perturbation, that is assumed to have the form of a plane
wave $\sim\exp (i\mathbf{k\cdot x}+\omega t)$, is imposed on an
initial equilibrium configuration. This procedure generally leads
to the formulation of the stability problem as a linear eigenvalue
equation, from which a dispersion relation
$\omega=\omega(\mathbf{k})$, which relates the wavenumber
$\mathbf{k}$ of the perturbation to its complex frequency
$\omega=\gamma+i\omega_i$, can be obtained. Normal mode stability
analysis reduces ultimately to the study of the sign of $\gamma$,
and the general statement is that the system is stable for small
perturbations when $\gamma\leq 0$
for every $\mathbf{k}$, and unstable otherwise.\\
However, it is now acknowledged that normal-mode analysis can
dramatically fail to predict the short-term behavior of a system
if the linear operator is non-normal, producing a result which is
only valid in the large time asymptotic limit \citep{schmid07}. A
non-normal operator $\mathbf{A}$ is one that does not commute with
its adjoint: $\mathbf{AA^*}\neq \mathbf{A^*A}$. If $\mathbf{A}$ is
a matrix this is equivalent to saying that it does not have a
complete set of orthogonal eigenvectors. A fundamental
characteristic of a system described by a non-normal operator is
the possible presence of \emph{transient growth}: an initial perturbation
can be amplified by a very large factor in a short-time period,
even if the system is stable, i.e. all the normal modes are
predicted to damp exponentially.
%This effect can be very easily understood for
%a rank 2 matrix, and we refer to Figure (2) of reference
%\citep{schmid07} for an effective and powerful visualization of
%how perturbations are amplified despite damping eigenvectors.\\
%It is then clear that to study the dynamical evolution and the stability of
%such systems the standard normal mode approach is not sufficient,
%not being able to capture transient dynamics of small amplitude perturbations.\\
Non-normal linear operators have been extensively studied in
hydrodynamics \citep{schmid07}, especially for shear flow
configurations. The role of non-normality has been also considered
in plasmas, for drift waves \citep{camargo98}, for the resistive
Alfven paradox \citep{borba94}, and for shear flows
\citep{chagelishvili96}, but so far only in the fluid plasma
scenario. In this paper we present the first investigation of
non-normal effects in a model of a stable plasma which includes
kinetic effects. We will show that the linear operator that
describes the evolution of the plasma is non-normal, and that the
non-normality is an intrinsic characteristic of the kinetic
treatment of the plasma. Transient growth is more accentuated for
smaller scales and higher plasma beta, that is when the MHD
description becomes more and more inadequate, and the plasma can
be correctly described only with the inclusion of kinetic effects.
Moreover, while previous studies focused on MHD drift
instabilities, we will study a Maxwellian plasma in a homogeneous
magnetic field, i.e. we will focus on a stable plasma. In this
case the departure from the evolution predicted by normal mode
analysis is most evident, since perturbations will grow instead of
decay for a certain period of time. A Maxwellian plasma in a
homogeneous magnetic field does not dispose of any source of free
energy. Therefore, the amplification of a small fluctuation is
energetically driven solely by the particular initial conditions
of the disturbance that, perturbing the initial equilibrium,
provides a small input of free energy to the system. The inclusion
of the short-time dynamics described in the following can
completely distort the physics of some phenomena where the
validity of the linear theory is generally accepted. For instance,
some recent models have been developed to understand the role of
kinetic waves for the small-scale dissipation of turbulent energy
in the solar wind \citep{howes08b}. Numerical simulations have
shown that the linear approximation is a valid ansatz in this
case, and therefore it is a scenario where the results of the
present Letter might be
applicable.\\
%\section{Mathematical model}
A full kinetic treatment of a plasma would require in principle
the use of the Vlasov-Maxwell equations. The standard linear
theory for those equations is usually formulated as an initial
value problem, and the dispersion relation is found via a
non-linear eigenvalue equation $\mathbf{D\cdot E}=0$ (with
$\mathbf{D}$ a $3\times 3$ complex matrix, and $\mathbf{E}$ the
amplitude of the perturbed electric field, assumed to vary as a
normal mode). The Vlasov-Maxwell equation is not yet amenable as a
linear eigenvalue problem, and for this reason we will use instead
the linear equations of the Landau fluid model \citep{sulem08}.
This is a hierarchical set of fluid equations for the dynamics of
both protons and electrons, neglecting electron inertia, truncated
at the fourth order moments, with a closure relation that
evaluates higher-order moments using linear kinetic theory. The
Landau fluid model includes linear Landau damping and finite
Larmor radius corrections, and it has been shown to correctly
reproduce the linear dynamics of Kinetic Alfven waves
\citep{passot07}, and of the mirror instability \citep{passot06}.
We have used a nonlinear Landau fluid code to confirm all the
predictions of the linear theory, including the non-normal effects
we describe here. We study the linear evolution of an
electron-proton Maxwellian plasma in a homogeneous magnetic field
$B_0$. The temperature and the density are chosen to be equal for
electrons and protons. We follow the formalism for the study of non-normal
operators given in the monograph by \emph{Trefethen and Embree} \citep{trefethen_book}. The
advantage of using a fluid model instead of a fully kinetic one is
that, in the version used here, it leads to a linear eigenvalue
problem with 16 equations for the following physical quantities:
density, velocity, magnetic field, pressure, heat flux, and the
gyrotropic part of the fourth-order cumulant tensor. The
coefficient matrix is therefore a $16\times 16$ (sparse) complex
matrix, which makes the problem computationally affordable without
the use of any particular method used for large matrix
eigenvalue problems. The entries of the matrix are generated with a symbolic algebra software, and the 16 linear equations of the model will not be reported here, due to lack of space. A detailed description of the model can be found in Ref. \citep{passot07} and references therein.\\
Although we do not make use of the normal mode ansatz, we use the
Fourier transform in space of all quantities:
$\Phi(x,t)=\Phi(t)\exp(i\mathbf{k\cdot x})$. The linear set of
equations can be formulated as $dy(t)/dt=\mathbf{A}y(t)$, where
$y$ is the state vector composed of the amplitude of the 16
variables, and the matrix $\mathbf{A}$ is a function of
$\mathbf{k}$, and of the plasma beta, $\beta=\frac{8\pi n
T}{B_0^2}$. The solution of the linear equation is
given by $y(t)=e^{\mathbf{A}t}y(0)$. We define $G(t)=\|y(t)\|/\|y(0)\|$,
where $\|\cdot\|$ is the euclidean 2-norm. The quantity
$G(t)$ gives the amplification (or reduction) of a
perturbation, i.e. the amplitude of that perturbation in time,
relative to its initial value: $G(t)=\|e^{\mathbf{A}t}y(0)\|/\|y(0)\|$. The supremum of $G(t)$ over all non-zero vectors $y(0)$
is the standard definition of the norm $\|e^{\mathbf{A}t}\|$. This quantity
defines the envelope curve which bounds from above the evolution of $G(t)$ for all possible
perturbations. Although in general the amplification of a
perturbation could stay well below $\|e^{\mathbf{A}t}\|$, and one single
perturbation will not reach maximum amplification for all times,
we here use the same assumption that is always implicitly made in
linear plasma theory. That is we assume that all the possible
perturbations of the system are excited, and we will focus on the
particular one that reaches the maximum possible amplification
$M=\max \|e^{\mathbf{A}t}\|$. We notice that the 2-norm of the state vector is not strictly related to the perturbed energy of the system, as it is usually done in works dealing with hydrodynamics nonmodal theory \citep{schmid07}. This is because in our model the state vector $y$ contains variables (such as high order moments) that do not enter in the expression for the energy. The norm of the state vector has therefore to be considered only as a measure of the perturbation applied to the system. It is obvious that a large (norm of the) perturbation implies a deviation from the assumption of linearity, which could result in the triggering of non-linear effects. In other words, transient growth of the euclidian norm of the state vector is physically equivalent to the development of an instability.\\
The key aspect of non-normal operators is that the spectrum may be highly sensitive to small perturbations. As a consequence it is
difficult, or rather improbable, for an initial perturbation to
excite only a single mode of the system. This is due to the
non-orthogonality of the eigenvectors: a state vector that
slightly deviates from lying on a single eigenvector can have
large projections on other eigenvectors, thus resulting in the
excitement of other modes. This is not the case if the
eigenvectors are all mutually orthogonal, as for normal operators.
A mathematical tool to characterize this behavior is given by
the concept of pseudospectrum, which is a generalization of the
standard spectrum. The spectrum $\sigma(\mathbf{A})$ is defined as
the set of points $z\in \mathbb{C}$ for which the resolvent matrix
$(z\mathbf{I}-\mathbf{A})^{-1}$ does not exist or, conventionally
$\|(z\mathbf{I}-\mathbf{A})^{-1}\|=\infty$. The
$\varepsilon$-pseudospectrum $\sigma_\varepsilon(\mathbf{A})$ of
$\mathbf{A}$ is the set of $z\in \mathbb{C}$ such that
$\|(z\mathbf{I}-\mathbf{A})^{-1}\|>\varepsilon^{-1}$ or,
equivalently, $z$ is an eigenvalue of the matrix $(\mathbf{A+E})$
for some matrix $\mathbf{E}$ with $\|\mathbf{E}\|<\varepsilon$
\citep{trefethen_book}. So the $\varepsilon$-pseudospectrum gives
a measure of how the spectrum is distorted due to a perturbation
of the operator of size $\varepsilon$. Physically one can think
that a perturbation on the evolution linear operator is equivalent
to perturbations or
inhomogeneities of quantities such as density or magnetic field.\\
In passing we note that the concept of pseudospectra has never
been emphasized in the analysis of numerical plasma simulations,
even though it is a common experience to see `transient effects'
at the beginning of simulations (which could also, of course, have
other causes). In particular, works that address the decay of a
single normal mode should be interpreted within this context,
since numerical fluctuations (especially in PIC codes where they
are unavoidable) approximately play the role of perturber of the linear operator.\\
Pseudospectra are a convenient graphical tool for understanding
the behavior of an operator. For a normal matrix, the
$\varepsilon$-pseudospectrum is just the union of the open
$\varepsilon$-balls about the point of the spectrum:
$\|(z\mathbf{I}-\mathbf{A})^{-1}\|=1/\text{dist}(z,\sigma(\mathbf{A}))$,
where $\text{dist}$ indicates the distance of a point to a set in
the complex plane \citep{trefethen_book}. We plot in Figure
(\ref{contour_inset}) an example of the contours of the
$\varepsilon$-pseudospectrum of our Landau fluid operator (not all
the eigenvalues are shown). The interpretation of the contours is
that a perturbation $\varepsilon$ will move the spectrum within
the region bounded by the $\varepsilon$-contour
($\varepsilon$-pseudospectra are nested sets, so that
$\sigma_{\varepsilon_1}(\mathbf{A})\subseteq\sigma_{\varepsilon_2}(\mathbf{A})$
for $\varepsilon_1\leq\varepsilon_2$). It is clear that small
perturbations make
some normal modes become connected to each other, and result in a distortion of the spectrum. For instance the contour for $\varepsilon=10^{-2.9}$ encloses all the 9 eigenvalues, which means that the system has completely lost the information about its exactly unperturbed solutions, since the solutions for the lightly perturbed system could lie anywhere within the $\varepsilon$-contour.
We also plot with a dotted line the $\varepsilon$-contour for $\varepsilon=0.5$, as it would be if the operator were normal.
In order to obtain the same kind of distortion of the spectrum for a normal operator, the perturbation has to be ~400 times larger.\\
The damping rate of the least damped mode $\alpha(\mathbf{A})=\max[\Re(\sigma(\mathbf{A}))]$ is the object
of the normal mode stability analysis, and dictates the behavior
at large times: $\lim_{t\to\infty} t^{-1}\log \|e^{\mathbf{A}t}\|=\alpha(\mathbf{A})$.
If the linear operator were normal this would be also the damping
rate of $\|e^{\mathbf{A}t}\|$ for any initial perturbation for any time $t\geq 0$, and there
would be no transient growth. In general however the initial
growth of $\|e^{\mathbf{A}t}\|$ is defined as the numerical abscissa
$\eta(\mathbf{A})=\frac{d}{dt}\left\vert \| e^{\mathbf{A}t}\| \right\vert_{t=0}$, which is given by the
formula \citep{trefethen_book}: $\eta(\mathbf{A})=\sup\sigma\left(\frac{1}{2}(\mathbf{A}+\mathbf{A}^*)\right)$, from which it is evident that $\eta(\mathbf{A})=\alpha(\mathbf{A})$ for a normal
matrix \footnote{Note that the numerical abscissa is denoted with $\omega(\mathbf{A})$ in reference \citep{trefethen_book}}.\\
The final definition we provide is the `departure from normality'
$D(\mathbf{A})$, which is a scalar that defines `how non-normal'
an operator is. There are several different way to characterize
$D(\mathbf{A})$, and we refer again to reference
\citep{trefethen_book} for more details. In the following we will
use the definition due to Henrici: $\mathbf{A}$ can be Schur
decomposed $\mathbf{A}=\mathbf{U(\Lambda+R)U^*}$, where
$\mathbf{U}$ is unitary, $\mathbf{\Lambda}$ is diagonal, and
$\mathbf{R}$ is strictly upper triangular. When $\mathbf{R}$ is
zero, $\mathbf{A}$ is normal, hence we define $D(\mathbf{A})=\|\mathbf{R}\|_F$, where $\|\mathbf{A}\|_F=\sqrt{\text{Tr}(\mathbf{AA^*})}$ is the Frobenius norm.\\
We present now a parametric study for a plasma subject to an
oblique perturbation with an angle $\theta=70^\circ$ between the
wavevector and the background magnetic field. We notice that the
Landau fluid model has a domain of validity that extends to small
scales only for oblique wavevectors. We span the range $[0.1, 10]$
for the values of $k$ and $\beta$. In Figure (\ref{num_abs-dep-k})
we show the numerical abscissa $\eta(\mathbf{A})$ (solid line),
and the departure from normality $D(\mathbf{A})$ (dashed line) as
functions of the wavenumber $k$. Here and in the following
figures, $k$ is normalized to the ion Larmor radius. The four
curves are for different values of $\beta=0.1,1,5,10$, from the
lower to the upper curve. All the curves are monotonically
increasing denoting that
the degree of non-normality increases with $k$ and $\beta$.\\
In order to understand how large the transients can grow, we show in Figure (\ref{sim}) the value of $G(t)$ for four different cases, with $\beta=1,10$, and $k=1,10$. Figure (\ref{sim}) was produced with values computed via a non-linear Landau fluid code, and the predictions from the linear theory (not shown) agree perfectly with the non-linear simulations (any differences would not be noticeable, if plotted on the same figure). The small initial perturbations have been chosen so that $\max G(t)=M$.
%As a general feature one can notice that an increase in $k$ for fixed $\beta$ moves the maximum amplification $M$ to
%smaller times, without changing dramatically its value, while an increase of $\beta$ for fixed $k$ not only shifts the peak to smaller times, but also increases its value.
We emphasize that the time scales involved in Figure (\ref{sim}) are a central point of our argument. One could indeed reasonably ignore all transient effects that take place on times much smaller than the typical timescales of the plasma. However one can see that, depending on the case considered, the amplification $G(t)$ can reach a value $\sim10^3$ for a time equal to $1/10$ of a ion gyroperiod, or even $10^4$ for $T\Omega_i\sim5\times10^{-2}$. Those are times scales where electron dynamics are important. Also, looking at the time $T\Omega_i=1$, all the four curves are within the interval $[1,100]$. This suggests that in some cases the protons also could be influenced by transient behavior.\\
The dependence on the values of $\beta$ and $k$ of the maximum
amplification $M$ and of the time $\tau$ at which this
amplification is reached cannot be easily inferred from Figure
(\ref{sim}). It turns out that while $\tau$ is a monotonic
decreasing function of $k$ for any $\beta$, $M$ changes its
behavior as $\beta$ varies. For small $\beta$, $M$ increases for
increasing $k$, while at high $\beta$ it decreases when $k$
increases. \\
%Interestingly, we have found that the following
%empirical relation holds (for the range of $\beta$ and $k$
%studied):
%\begin{equation}\label{fit}
%M/\tau = m_1k^{m_2}\beta^{m_3}
%\end{equation}
%with fitting parameters $m_1\sim 6600$, $m_2\sim 0.95$, $m_3\sim 1.5$. Although Eq. (\ref{fit}) does not provide the individual values for $M$ and $\tau$, it gives the physical insight that for large $k$ and $\beta$ large transient growth can survive only for short times.\\
Having seen that transient growth can reach large values of amplification we address now the point of how long the transient effect can last. We plot in Figure (\ref{contour}) contours of the time $\Theta$ for which $\|e^{\mathbf{A}\Theta}\|\leq 1$, as a function of $k$ and $\beta$. This is the time when $G<1$ for any initial perturbation and we consider it as the time when transient effects lose importance, and the system starts to damp according to the normal-mode analysis. It appears that transient growth takes a longer time to disappear at small $k$. Also, there is a large portion of parameter space where $10<\Theta\Omega_i<100$, and therefore the transient time is comparable with typical plasma time scales.\\
We have shown that a collisionless plasma modeled through a Landau-fluid model is a non-normal system where transient growth can take place for short periods of time. By studying how the transients behave as functions of $k$ and $\beta$, we argue that the non-normality of the system is connected with small scales and high plasma $\beta$. For this reason, although one is not permitted to generalize \emph{tout court} the results of the present Letter to a Vlasov plasma, we conjecture that the Vlasov equation should also present non-normal effects. Hence, the key point is to understand if the transients can affect the plasma dynamics, and therefore whether a generalized nonmodal plasma instability theory should supersede in some cases the more traditional normal-mode analysis. This question is intuitively related to how large the transients can grow, and how long they last. We have shown that, for the Landau-fluid model, the transients are surprisingly large, with possible amplification of the order of $10^3-10^4$; they can also last impressively long, for times of 10-100 ion gyroperiods.\\
We believe that the results of the present Letter will have important and immediate applications in at least two areas. The first is the study of kinetic turbulence in space plasma, where the linear approximation is universally accepted and where models have so far neglected any non-modal prediction. The second deals with the interpretation of computer simulations, especially those that study damping waves in a stable plasma. In the light of these results, the idea that a single normal mode can be numerically excited without triggering transient effects should be revisited.\\
To conclude, the results presented here point towards a revision of kinetic plasma theory from a non-modal perspective: a route that has already successfully been followed in hydrodynamics.

\newpage

\begin{acknowledgments}
We acknowledge Dr. Oscar Bandtlow and Dr. Gian Luca Delzanno for useful comments.
\end{acknowledgments}


%\bibliography{../camporeale_bib}

\end{document}